# Quantifying classical entanglement using polarimetry: Spatially-inhomogeneously polarized beams

**C.T. Samlan and Nirmal K. Viswanathan**[*]

School of Physics, University of Hyderabad, Hyderabad – 500046, India

[*]Email: nirmalsp@uohyd.ernet.in

Generation of spatially-inhomogeneously polarized (SIP) beams due to polarization – spatial mode non-separability is demonstrated using a polarization-Sagnac interferometer. Counter-propagating horizontal ($H$) and vertical ($V$) polarized Laguerre-Gaussian beams of opposite topological charges ($LG_0^{\pm 1}$), corresponding to the two binary degrees of freedom (DoF) are coherently superposed in the interferometer. Quantum-inspired entanglement witness and quantification such as Bell state measurement, violation of CHSH form of Bell's inequality ($B = 2.33 \pm 0.024$) and concurrence measurements ($C = 0.912 \pm 0.027$) are experimentally measured to establish non-separability between the two DoF. As the SIP beams are equivalent to bipartite pure-states, point-wise Stokes polarimetry measurements are used to calculate the reduced coherency matrix corresponding to polarization DoF. Using these we calculate the linear entropy ($S_L = 0.968 \pm 0.031$) and maximized Bell's measure ($B_{\max} = 2.824 \pm 0.004$) as to quantify the degree of entanglement of the SIP beams.

## I. INTRODUCTION

Entanglement until recently is been quintessentially non-classical, wherein the existence of nonlocal correlation in quantum systems such as trapped ions, spin qubit in semiconductors and quantum dots and photon qubits, are utilized to address important issues in the development of foundations and applications of quantum mechanics, quantum optics and quantum information. More recently, quantum-like non-separable structures in optical beams, due to correlation between two degrees of freedom (DoF) of classical light fields, has been demonstrated in deterministic 'spin-orbit' [1] and non-deterministic Shimony-Wolf [2] states of light, in addition to such demonstrations in light beams with topological singularity [3] and in vector-vortex beams [4]. These results have provided significant impetus to the emerging area of research on 'classical entanglement' or non-separability between two DoF in optical beams, which has also been identified as a resource to resolve the long-standing issue concerning Mueller matrices [5] and its application to polarization metrology [6], in the unification of competing interpretations of degree of polarization [7] and in the application of Bell's measure as a new index of coherence in optics [8]. With the recent realization that cylindrical vector beams (CVBs) are a non-separable combination of polarization-spatial mode DoF [4, 9] the interest in this area of research is only bound to increase.

At the outset we would like to differentiate classical entanglement (CE) from quantum entanglement (QE) [10] by emphasizing that in CE (i) there is *no non-locality*, as the entanglement occurs between different attributes (polarization and spatial mode, for example) of the same light beam; (ii) the optical modes of interest are the ones with large number of photons, and so there is *no measurement-induced collapse* due to photon indivisibility, and hence there are *no mutually exclusive outcomes*. Thus, there is no definite relation between classical and quantum entanglement. Nevertheless, the similarities between these two domains of entanglement are important, since the idea of superposition of states in Hilbert space is shared by both quantum physics and classical wave physics, and because of the mathematical isomorphism of correlations between discrete degrees of freedom in classical optics with QE in two-qubit bipartite systems [11-13]. It is thus worth emphasizing that the Hilbert space structure of classical optics enables the realization of entanglement, in the sense of non-separability of the polarization and spatial mode DoF of a classical optical beam. These similarities benefited the implementation of classical optics tools to quantum computational tasks (without nonlocality requirement) [14], such as in Grover and Deutsch algorithms [15, 16], quantum walks using orbital angular momentum [17], realization of entangled polarization and spatial modes of classical light as contextual variables [18, 19], and to open new avenues for (quantum) computation and communication tasks [20] using classical entanglement as a resource [6, 9].

Cylindrical vector beams (CVBs) have been known for a long time [21] but it was pointed out only recently that these beams are non-separable states of light, due to the direct product of polarization and spatial mode vector spaces [4, 6, 9, 13]. As suggested, the occurrence of entanglement requires a mathematical expression describing the whole beam to be written as the sum of tensor product of two or more vectors, belonging to different vector spaces [6]. The non-separable superposition of polarization vector space (represented by Poincaré sphere) [22] and the first-order spatial modes (represented by spatial mode sphere) [23] results in CVBs which can be represented on a hybrid Poincaré sphere (HPS). The optical beams thus formed with hybrid degree of freedom of the electromagnetic field describe neither purely polarized nor pure spatial modes of the field, and are thus considered entangled [4, 24]. For example, radially polarized optical beam with zero net angular momentum, formed due to the superposition of horizontal $|H\rangle$ and vertical $|V\rangle$ polarized Hermite-Gaussian modes ($\psi_{10}(r), \psi_{01}(r)$) can be written as $E(r) = 1/\sqrt{2}\left[|H\rangle\psi_{10}(r) + |V\rangle\psi_{01}(r)\right]$, or in an isomorphic form of two qubit Bell state as $|E\rangle = 1/\sqrt{2}\left[|0,0\rangle + |1,1\rangle\right]$ [6]. It is important to note here that the CVBs are a special class of spatially inhomogeneously polarized (SIP) beams that can be generated by a coherent superposition of any pair of orthogonal states in polarization and spatial modes. Once generated, a SIP beam can be transformed to other states on the HPS via unitary transformations, using waveplates and / or mode converters, without affecting the degree of entanglement, as will be shown in this article.

The non-separability between different DoFs in optical beams, also known as hybrid entanglement is established and quantified using different quantum-inspired measurements: the indicator for entanglement, violation of Bell's inequality provides strong evidence of non-separability between polarization and spatial mode DoF; concurrence and entropy measurements provide means to quantify entanglement between the DoF. For a bipartite quantum system in pure state, Bennett et al., [25] have shown that von Neumann entropy of either of the two DoF is a reasonable measure to quantify entanglement of the system: if $|\psi_{AB}\rangle$ is the state of the bipartite system, the amount of entanglement is given by its Neumann entropy, $S(\psi) = -Tr(\rho_A \log_2 \rho_A) = -Tr(\rho_B \log_2 \rho_B)$, where $\rho_{A(B)} = Tr_{B(A)}\rho$ is the reduced density matrix of each subsystem and $\rho = |\psi_{AB}\rangle\langle\psi_{AB}|$ [26]. The amount of entanglement of a bipartite pure state is considered optimal when the state projections are measured along mutually

orthogonal and linearly independent directions [26]. In addition we show in this article that for the SIP beams represented on the HPS, projecting one of the DoF (state of polarization − SoP, for example) in different angular orientations directly influences the angular distribution of the other DoF (spatial mode): different SoP projections of the beam with non-separable DoF results in different mode projections and vice-versa due to strong correlation between the two DoF. Expressed in terms of the Bell state, the fully characterized SIP beam has no information about each of the subsystems used to create the non-separable beam.

The article presents the generation of SIP beam due to non-separable polarization − spatial mode DoF and quantum-inspired Bell-state measurement, violation of Clauser, Horne, Shimony and Holt (CHSH) form of Bell's inequality and concurrence as entanglement witness and to quantify entanglement. In addition, reduced coherency matrix corresponding to polarization DoF is used to calculate linear entropy and maximized Bell's measure to quantify the degree of entanglement. These are described and discussed in different sections: theoretical details necessary to understand the generation of SIP beams as due to non-separability between polarization and spatial mode DoF is given in section II A. Details necessary to understand and quantify the degree of entanglement of the SIP beams using polarimetry measurements, beginning with an analogy with bipartite pure state quantum entanglement is given in section II B. Experimental details discussing the salient features of the polarization Sagnac interferometer (PSI) used to generate the SIP beams and classical methods used to analyse the beam are given in Section III. The non-separability between the two DoF is established via Bell state measurement, violation of the CHSH form of Bell's inequality and concurrence measurements. As the SIP beams are considered equivalent to bipartite pure quantum states, reduced coherency matrix corresponding to polarization DoF is calculated using point-wise Stokes polarimetry measurements. Using these we calculate the linear entropy and maximized Bell's measure to quantify the degree of entanglement of the SIP beams. Continuous variation of the superposed beams' SoP takes the SIP beam from the condition of maximal non-separability to beams with separable DoF. In addition, by passing the SIP beams through waveplates we transform between the different CVBs. All the experimental results obtained are presented, analyzed and discussed in section IV and the work is finally summarized in section V.

## II. THEORETICAL DETAILS

### A. Spatially inhomogeneously polarized beams

Optical beams wherein the SoP varies around the axis in a cylindrically symmetric way belong to the extensively researched CVBs [21, 27]. These axially symmetric CVBs are a special class of the generic vector-vortex beam-fields, also known as SIP beams, wherein the state of polarization (SoP) varies in general in the beam cross-section and which are a solution to the full vector electromagnetic wave equation. In general, the SIP beams can have non-zero total angular momentum / Poynting vector density [30], contrasting with radial, azimuthal and hybrid polarized CVBs with zero total angular momentum [31]. The intense interest in these beams, apart from being of fundamental interest is due to the immense potential such beams offer in a number of emerging applications in microscopy-nanoscopy, imaging, information encoding etc. [21, 27 – 29].

The CVBs are understood as a superposition of two-dimensional polarization $\{\hat{e}_x, \hat{e}_y\}$ and spatial mode $\{\psi_{10}, \psi_{01}\}$ spaces [31]. Using the Poincaré sphere representation for fully polarized optical beams and an analogous Bloch mode sphere representation for coherent OAM modes, the CVBs have been represented on the hybrid Poincaré sphere (HPS) [31, 32]. Each point on the HPS can be characterised completely by the three Stokes or equivalent parameters. Such a representation of the classical optical beam can be considered equivalent to the bipartite pure state of a quantum system represented on the Bloch sphere, as the beam characteristics are completely specified and identified on the surface of the corresponding sphere representation. Thus the cylindrically polarized states of light can be visualized in terms of hybrid Stokes parameters and transformed on the HPS [31]. However, to perform measurements to represent the CVB on the HPS an arbitrary cylindrically polarized state of light is to be projected in the Stokes polarization and modal basis of interest. This requires elaborate (and often difficult to manage) experimental setup consisting of asymmetric Mach-Zehnder interferometers and polarization optical elements, as shown in Fig. 6 of Ref. [31], unless the CVBs are considered analogous to maximally entangled two qubit Bell states [6, 9, 24].

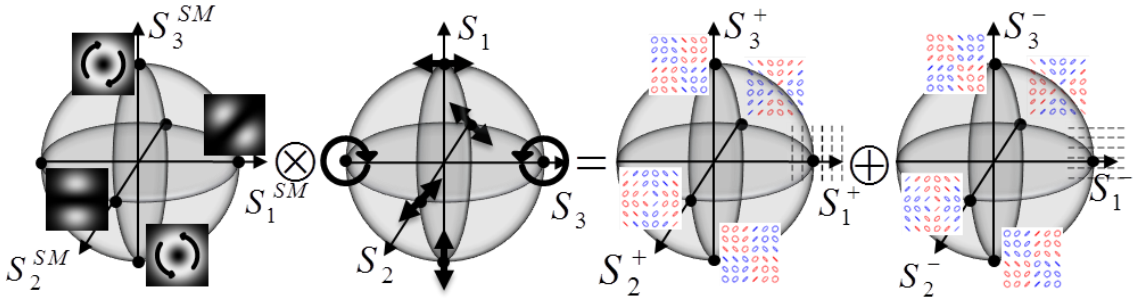

FIG.1. (Color online) Tensor product of spatial modes represented on modal sphere and state of polarization on the Poincaré sphere results in Cartesian sum of hybrid Poincaré sphere (HPS) representing entangled '+' and '-' modes. Two poles of both HPS ($+S_3^\pm$ and $-S_3^\pm$) represent the four Bell-states of non-zero total angular momentum and $S_1^\pm$ separable states of the beam, respectively.

Now consider the superposition of orthogonal linearly polarized Laguerre-Gaussian ($LG_0^{\pm 1}$) modes with $l = \pm 1$ units of orbital angular momentum (OAM). The horizontal ($\hat{e}_H$) and vertical ($\hat{e}_V$) polarized first-order LG modes, designated as $\psi_l$ and $\psi_r$ respectively for clock-wise and counter clock-wise rotating paraxial vortex modes, carry $\pm 1$ units of OAM [33]. We follow here the notations and treatment used in Refs [6, 31] for designating the spatial modes and their SoP to construct the HPS representation for the resulting SIP beams. As shown in Fig. 1, representing the SoP on the Poincaré sphere and the spatial modes on the analogous modal sphere, the coherent superposition of orthogonal linearly polarized first-order LG modes can be represented as the Cartesian sum of two HPS that embody all characteristics of the complex modes of the electromagnetic field. The north and south poles of both the HPSs, $+S_3^\pm$ and $-S_3^\pm$ represent the four Bell states $\Phi^\pm = \frac{1}{\sqrt{2}}(|\hat{e}_H \psi_l\rangle \pm |\hat{e}_V \psi_r\rangle)$ and $\Psi^\pm = \frac{1}{\sqrt{2}}(|\hat{e}_H \psi_r\rangle \pm |\hat{e}_V \psi_l\rangle)$ with non-zero total angular momentum and $S_1^\pm$ correspond to separable states of the beam. The spatially varying polarization state of the Bell state beam due to non-separable polarization and spatial mode combinations is understood by decomposing the input beam into its polarization and spatial mode components and superposing them as shown in Fig. 2 to realize the $\Phi^+$ Bell state of the SIP beam. This understanding is useful in the Bell state measurement discussed in the experimental section.

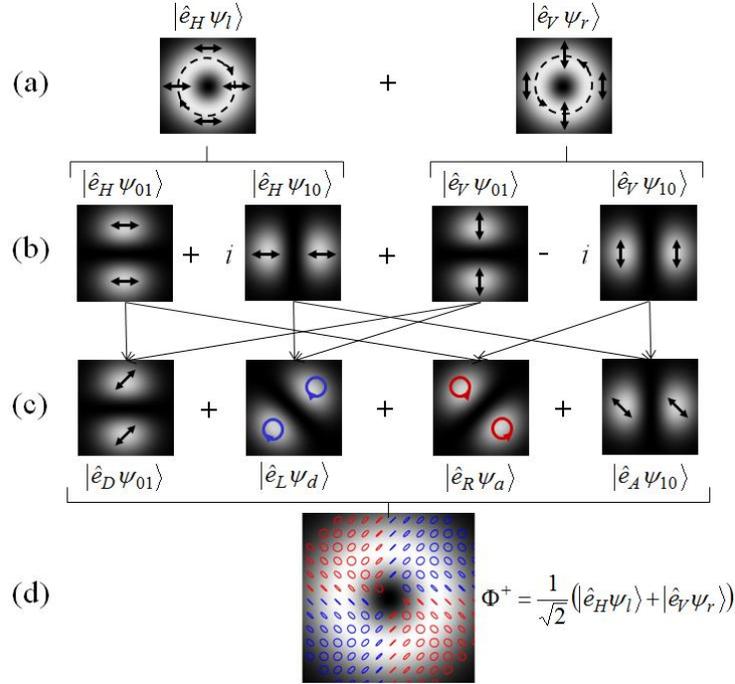

FIG. 2. (Color online) Schematic to understand the formation of $\Phi^+$ Bell state beam by decomposing the input linearly polarized LG modes (a) into its components (b), which are superposed (c) to form the SIP beam (d).

Though the four Bell states do not belong to radial, azimuthal or hybrid CVBs with zero total angular momentum, by passing the SIP beams through suitable wave plates or mode converters (unitary transformation) the beam can be changed to any of the CVBs (and be represented on the HPS). The Bell state of interest to us here can be written in a generalized form as $\Phi^+ = \cos\theta |\hat{e}_H \psi_l\rangle + \sin\theta e^{i\phi} |\hat{e}_V \psi_r\rangle$ where, $\phi$ is an arbitrary phase that is included to match with the experimental results and $\theta$ is the intensity weightage of the fields that can act as a tunable parameter of the hybrid entanglement: one of the maximally non-separable (MNS) Bell state $\Phi^+$ is realized for $\theta = \pi/4$, and separable states when $\theta = 0$ or $\pi/2$. For all other values of $\theta$ partially entangled SIP beam is obtained. Alternately, the SIP beam wherein the SoP varies azimuthally can also be understood via the generalized interference equation given in terms of Stokes parameters $(S_i; i = 0-3)$ [34]

$$S_i(\bar{r},\varphi) = S_i^l \left(\cos\theta |\psi_l(\bar{r},\varphi)\rangle\right)^2 + S_i^r \left(\sin\theta |\psi_r(\bar{r},\varphi)\rangle\right)^2 + \sin(2\theta)|\psi_l(\bar{r},\varphi)\rangle|\psi_r(\bar{r},\varphi)\rangle \operatorname{Re}\left[e^{2i\varphi}\varepsilon_l^* \sigma_i \varepsilon_r\right]$$

and the Stokes interference term corresponding to the $\Phi^+$ Bell state is given by

$$\vec{S}(\bar{r},\varphi) = \begin{bmatrix} \left(\cos\theta|\psi_l(\bar{r},\varphi)\rangle\right)^2 + \left(\sin\theta|\psi_r(\bar{r},\varphi)\rangle\right)^2 \\ \left(\cos\theta|\psi_l(\bar{r},\varphi)\rangle\right)^2 - \left(\sin\theta|\psi_r(\bar{r},\varphi)\rangle\right)^2 \\ \cos 2\varphi \sin 2\theta |\psi_l(\bar{r},\varphi)\rangle|\psi_r(\bar{r},\varphi)\rangle \\ -\sin 2\varphi \sin 2\theta |\psi_l(\bar{r},\varphi)\rangle|\psi_r(\bar{r},\varphi)\rangle \end{bmatrix} \tag{1}$$

Equation (1) clearly shows modulation in Stokes parameters $S_2$ and $S_3$ depending only on the azimuthal angle $(\varphi)$ and an absence of modulation in $S_0$ and $S_1$, as shown in Fig. 2 (d). It is understandable that interference between orthogonal polarizations do not produce any intensity modulation, corresponding to vanishing $S_0$, when $\psi_l$ and $\psi_r$ intensities are equal and generation of one of the Bell states when $\theta = \pi/4$ with absence of $S_1$, as in our case.

### B. Polarimetric measurement of degree of entanglement

Consider bipartite pure quantum state $|\psi_{AB}\rangle$, a vector in the 4D Hilbert space $H_{AB} = H_A \otimes H_B$, where $H_A$ and $H_B$ are the 2D Hilbert spaces corresponding to its two subsystems $A$ and $B$. The density matrix of the state $|\psi_{AB}\rangle$ is given by $\rho_{AB} = |\psi_{AB}\rangle\langle\psi_{AB}|$ and the reduced density matrix corresponding to the subsystem $A$ is,

$$\rho_A = Tr_B(\rho_{AB}) \tag{2}$$

Where $Tr_B$ is the partial trace defined as $Tr_B(\rho_{AB}) = \sum_{n=1}^{2}\langle \hat{e}_n^B | \rho_{AB} | \hat{e}_n^B \rangle$ [35], where $\hat{e}_i$ are the bases of 2D Hilbert space $H_B$. Thus, the measurement of entanglement of the bipartite pure state is possible by measuring the reduced density matrix corresponding to either one of the subsystems [25, 26]. We use this analogy to quantify the non-separability between polarization and spatial mode DoF of the SIP beam. Projecting different SoP represented on the Poincaré sphere will result in different mode characteristics represented on the modal sphere and vice-versa. The point-wise Stokes polarimetry measurements are used to calculate the reduced coherence matrix corresponding to the polarization DoF, which is isomorphic to the reduced density matrix formalism. Thus by characterizing the polarization coherency matrix via the degree of polarization coherence, we quantify the degree of entanglement by calculating the linear entropy and maximized Bell's measure.

As discussed in Ref. 8, consider the 4x4 global coherency matrix, which is mathematically isomorphic to the density matrix formulation of bipartite state, that provides a complete description of the electromagnetic beam-field,

$$G = \langle E_i(r_m) E_j^*(r_n) \rangle_{i,j=x,y; m,n=1,2} \tag{3}$$

where, $r_1$ and $r_2$ are the two spatial points in the beam cross section. The reduced 2x2 coherency matrices that characterize the polarization and spatial mode DoF can be obtained from the global coherency matrix as

$$G_{pol} = \begin{pmatrix} \langle E_x(r_1)E_x^*(r_1)\rangle + \langle E_x(r_2)E_x^*(r_2)\rangle & \langle E_x(r_1)E_y^*(r_1)\rangle + \langle E_x(r_2)E_y^*(r_2)\rangle \\ \langle E_y(r_1)E_x^*(r_1)\rangle + \langle E_y(r_2)E_x^*(r_2)\rangle & \langle E_y(r_1)E_y^*(r_1)\rangle + \langle E_y(r_2)E_y^*(r_2)\rangle \end{pmatrix} \quad (4)$$

$$G_{sp} = \begin{pmatrix} \langle E_x(r_1)E_x^*(r_1)\rangle + \langle E_y(r_1)E_y^*(r_1)\rangle & \langle E_x(r_1)E_x^*(r_2)\rangle + \langle E_y(r_1)E_y^*(r_2)\rangle \\ \langle E_x(r_2)E_x^*(r_1)\rangle + \langle E_y(r_2)E_y^*(r_1)\rangle & \langle E_x(r_2)E_x^*(r_2)\rangle + \langle E_y(r_2)E_y^*(r_2)\rangle \end{pmatrix} \quad (5)$$

The polarization independent spatial mode coherency matrix $G_{sp}$ describes the correlation between the mode fields at two spatial points $r_1$ and $r_2$ while ignoring the polarization, whereas the $G_{pol}$ matrix describes the partial polarization of the field obtained by spatially averaging over the modes. This is in contrast to the usual polarization coherency matrix that describes a single spatial point in the beam field after integrating over space. The Eigen values of these reduced coherency matrices ($\mu_1^2$ and $\mu_2^2$) characterizes the degree of polarization (spatial) coherence given by $D_{pol} = D_{sp} = |\mu_1^2 - \mu_2^2|$, which can be calculated using the formula,

$$D_{pol(sp)} = \sqrt{1 - \frac{4\det\{G_{pol(sp)}\}}{\left(Tr\{G_{pol(sp)}\}\right)^2}} \quad (6)$$

Where, '$Tr$' and '$\det$' denote the matrix trace and determinant respectively. The reduced polarization and spatial mode coherency matrices are identical in this basis, $G_{pol} = G_{sp} = \begin{pmatrix} \mu_1^2 & 0 \\ 0 & \mu_2^2 \end{pmatrix}$. Reduced polarization coherency matrix can be obtained from the Stokes parameter measurements using the following relation [36]

$$G_{pol} = \frac{1}{2}\begin{pmatrix} S_0(r_1)+S_1(r_1) & S_2(r_1)-iS_3(r_1) \\ S_2(r_1)+iS_3(r_1) & S_0(r_1)-S_1(r_1) \end{pmatrix} + \frac{1}{2}\begin{pmatrix} S_0(r_2)+S_1(r_2) & S_2(r_2)-iS_3(r_2) \\ S_2(r_2)+iS_3(r_2) & S_0(r_2)-S_1(r_2) \end{pmatrix}$$

$$= \frac{1}{2}\left(\sum_{i=0}^{3} S_i(r_1)\sigma_i + \sum_{i=0}^{3} S_i(r_2)\sigma_i\right) \quad (7)$$

Where $S_{i=0-3}(r)$ are the normalized Stokes vectors at the spatial point $r$ and $\sigma_{i=1-3}$ are the Pauli spin matrices with $\sigma_0$ the identity matrix of rank two.

Entropy plays a significant role in both classical and quantum information theories, which are characterized via standard and mutually, related entropy measures – Shannon entropy and von Neumann entropy [37]. It has been established that the measures of entropy have been widely studied both in the context of classical probability distributions, and in pure and mixed quantum states [38], though for quantum systems, the easier to calculate linear entropy has been widely used [12]. For the classical SIP beams the linear entropy ($S_L$) [10] can be considered to quantify the degree of entanglement of fully polarized coherent beam which can be calculated from the degree of polarization coherence by using the complementary relation:

$$D_{pol}^2 + S_L^2 = 1 \quad (8)$$

For the SIP beam with maximally entangled DoF $S_L = 1$ and it is zero for the beam with separable DoF. The degree of entanglement is also calculated from maximized Bell's measure $(B_{max})$ calculated from the complementary relation:

$$4D_{pol}^2 + B_{max}^2 = 8 \qquad (9)$$

where, $B_{max}$ [8] varies from $2$ to $2\sqrt{2}$, corresponding to optical beams with separable and maximally entangled DoF.

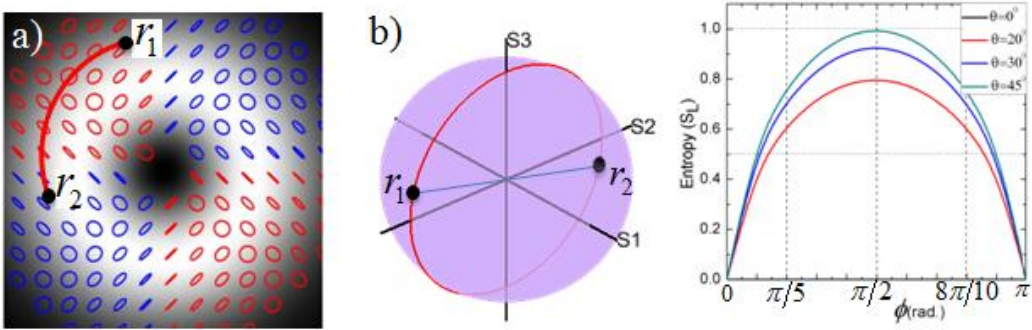

FIG. 3. (Color online) (a) Spatial points $r_1$ and $r_2$ are marked on a simulated maximally entangled beam cross section and red line denotes the integral path of geodesic; (b) The geodesic is plotted (red line) on the Poincaré sphere which is a great circle and the antipodal points (black dots) represent orthogonal polarization states corresponding to local points $r_1$ and $r_2$; (c) Simulated linear entropy for separable beam ($\theta = 0°$) and maximally ($\theta = 45°$), and partially ($\theta = 20°, 30°$) entangled beams calculated for different azimuthal separation ($\varphi$) of the two spatial points.

However, these complimentary relations (eqns. 8 and 9) hold true only when $D_{pol}$ is obtained from uniquely calculated $G_{pol}$ for the SIP beam. For which, it is possible to find pairs of spatial points $r_1$ and $r_2$ over the cross-section of the entire beam where the degree of polarization-coherency is minimum, equivalent to maximum degree of entanglement, which is similar to the maximum violation of the CHSH form of Bell's inequality over all the settings [3, 8]. In other words, for every point $r_1$ in the beam cross-section there is a corresponding point $r_2$ which gives reduced polarization coherency matrix for the entire beam and hence maximizes the entropy. The identification of the unique pair of points is carried out by fixing $r_1(r, \phi_0)$ and varying $r_2(r, \phi_0 + \varphi)$ on the beam cross section and calculating the linear entropy. The azimuthally separated spatial points $r_1$ and $r_2$ are shown on the SIP beam cross-section in Fig. 3 (a). The spatial points can be represented by mapping the local polarization states in the beam cross section along the polarization gradient (which is azimuthal direction) on the surface of the Poincaré sphere leading to the geodesic $(\Delta S)$. The Geodesic is defined by,

$$\Delta S = \int_{\phi_0}^{\phi_0 + \phi} \sqrt{\left(\frac{dS_1}{d\phi}\right)^2 + \left(\frac{dS_2}{d\phi}\right)^2 + \left(\frac{dS_3}{d\phi}\right)^2} \, d\phi \qquad (10)$$

Where, $S_{1,2,3}$ are the normalized Stokes parameters and $\phi$ is the azimuthal angle in polar coordinate. The antipodal points of the geodesic corresponding to the two spatial points $r_1$ and $r_2$ in the beam cross section gives a unique reduced coherence matrix, $G_{pol}$ of the polarization DoF and the geodesic will be the great circle for maximally-entangled beam (Fig. 3b). Figure 3 (c) shows the relation between the linear entropy and azimuthal separation ($\varphi$) between the points $r_1$ and $r_2$ for different simulated beams: the entropy is zero for beams with separable DoF, corresponding to $\theta = 0°$, is maximized for $\varphi = \pi/2$ corresponding to maximally-entangled beam for $\theta = 45°$ and is partially entangled for $\theta = 20°, 30°$ beams as shown.

We thus have a useful formalism to measure and quantify the degree of entanglement via linear entropy ($S_L$) and maximized Bell's measure ($B_{max}$) using eqns. (8) and (9) from point-wise Stokes polarimetry measurements and reduced coherence matrix of the polarization DoF for a fully polarized coherent beam, an equivalent to the formalism used in bipartite pure quantum state entanglement measurements.

### III. EXPERIMENTAL DETAILS

To experimentally realize the SIP beam of the form given by one of the Bell states, $\Phi^+ = \cos(\theta)|H\rangle|\psi_l\rangle + \sin(\theta)e^{i\phi}|V\rangle|\psi_r\rangle; \theta = \pi/4$, we construct a polarization Sagnac interferometer (PSI) shown in Figure 4. The interferometer is used to entangle horizontal ($|H\rangle$) polarized $LG_0^{+1}:|\psi_l\rangle$ spatial mode and vertical ($|V\rangle$) polarized $LG_0^{-1}:|\psi_r\rangle$ mode. The intensity-frequency stabilized He-Ne laser ($\lambda = 632.8$ nm; Newport, USA) is linearly polarized using a Glan-Thompson polarizer. The beam is passed through a half-wave plate (HWP) – quarter-wave plate (QWP) combination to suitably transform the input beam SoP to ensure the intensity of the two interfering beams is the same, and also to control the phase difference between the beams. This beam is incident on the polarizing beam splitter (PBS), which transmits H-polarization and reflects V-polarization, and they propagate in opposite directions within the interferometer, and exit through the other port of the PBS as shown in Fig. 4. A spiral phase plate (VPP, RP Photonics, USA) with unit charge at the laser wavelength is kept within the interferometer, such that, the counter propagating $|H\rangle$ and $|V\rangle$ polarized Gaussian beams are converted into LG beams and with +1 ($|\psi_l\rangle$) and -1 ($|\psi_r\rangle$) units of OAM respectively. These two orthogonal linearly polarized counter-rotating LG modes then interfere with each other upon returning to the PBS, and the superposed beam exits through the complementary port of the interferometer. Also, as the even number of total reflections is the same for both the beams after passing through the SPP, the OAM sign of the two interfering beams remain unchanged upon returning to the PBS. The PSI is a very stable configuration and highly suitable for measurements over a long period of time. Apart from the inconsequential overall phase factor $\phi$ between the two output polarization components, the setup is independent of the interferometer path lengths. By independently

changing the polarization and OAM modes by introducing a HWP or a Dove prism within the interferometer it is possible to realize all the four Bell states for any subsequent measurement.

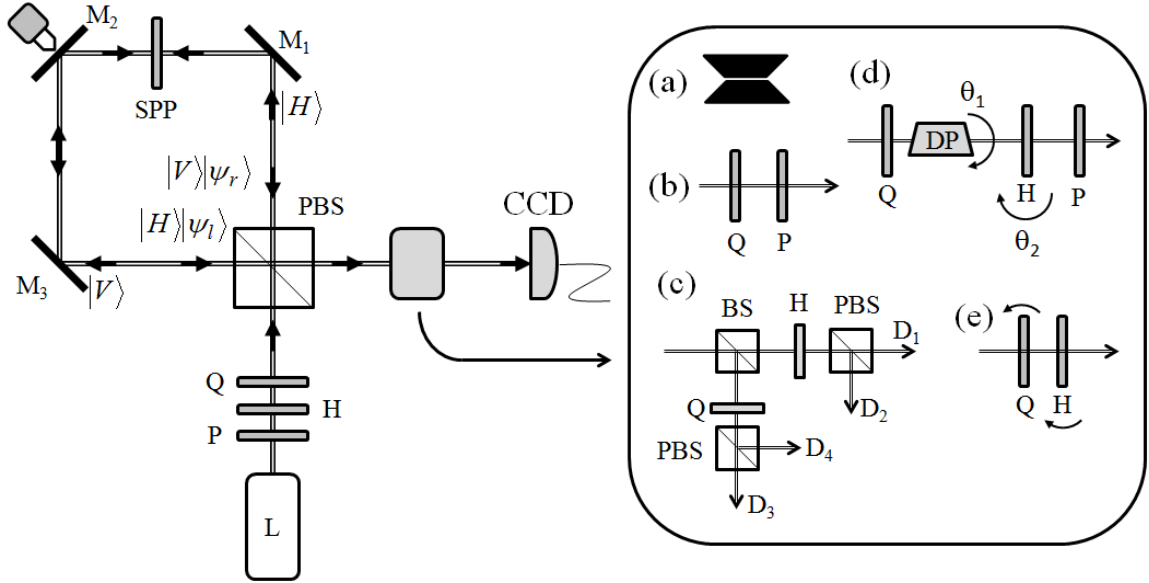

FIG. 4. Schematic of the polarization Sagnac interferometer (PSI) used to generate hybrid entangled optical beams. Inset shows the different setups at the interferometer output to measure (a) the phase distribution using a single slit diffraction, (b) Stokes parameters (with Q = 0°, 90°; P = 0°, 90°, 45° and 135°), (c) Bell state projections (with H = 22.5° and Q = 45°), (d) concurrence (with Q = 45° and P = 0°) and (e) rotatable Q and H waveplates to change between different CVB states. L – He-Ne laser, PBS – polarizing beam splitter, M – mirrors, SPP – spiral phase plate, Q (H) – quarter (half) wave plate, P – polarizer, DP – Dove prism.

The topological phase of the LG spatial mode is one of the important characteristic that can indicate the non-separability between the two modes of the PSI [39, 40]. Either a separate two-beam interferometer [20] or a single slit diffraction [41] experiment can be used to measure the topological phase of the output beam. Due to its simplicity we used a single slit diffraction experiment (Fig. 4 (a)) to comment on the topological phase of the output beam and its constituent beams from the interferometer, using the intensity pattern observed on the CCD camera. The non-separable DoF of the output beam is also evidenced by spatially non-uniform SoP in the beam cross-section and is measured using point-wise Stokes polarimetry [42]. This is carried out by passing the output beam and the constituent beams from the interferometer through a QWP-polarizer combination and measuring the Stokes parameters using a CCD camera (Fig. 4 (b)). The measurement details are given in our earlier publications [42]: briefly, six intensity measurements are made for different QWP-polarizer settings are the four Stokes parameters are calculated using which the point-wise polarization ellipse parameters are calculated. These measurements also enable us to construct the reduced polarization coherency matrix $G_{pol}$ using eqn. 7 and the degree of polarization coherence $D_{pol}$ using eqn. (6). From these the Bell's measure is maximized by considering two points $r_1$ and $r_2$ on the beam cross section such that $r_1 = r(\cos\phi + \sin\phi)$ and $r_2 = r[\cos(\phi + \pi/2) + \sin(\phi + \pi/2)]$. As mentioned in the theory section these points are identified by maximising the entropy or equivalently identifying the longest geodesic $(\Delta S = \pi)$ on the Poincaré sphere (eqn. 10). This is carried out by considering polarization gradient in the azimuthal direction and maximising it for all pairs of points $(r_1, r_2)$ in the beam cross-section. Equations (8) and (9) are then used to calculate the linear entropy ($S_L$) and maximized Bell's measure ($B_{max}$) of the maximally non-separable SIP beam generated using the PSI. Thus point-wise polarization measurements and subsequent calculations are used to quantify the degree of entanglement of the SIP $\Phi^+$ Bell state beam and for different CVBs, realized by making use of the quarter-half (QH) wave plate combination, introduced in the beam output from the PSI.

## IV. RESULTS AND DISCUSSION

The polarization Sagnac interferometer shown in Fig.4 is arranged such that the output beam is a coherent superposition of horizontal and vertical polarized $LG_0^{+1}$ and $LG_0^{-1}$ beams respectively. Though the two counter-propagating beams of same amplitude are expected to superpose in phase, a small (~ λ) non-zero phase difference appears, which is equalized by using the input QWP-HWP and / or by adjusting the mirror (M$_2$) mounted on a translation stage with fine tuning capability. An almost zero phase difference between the interfering beams is ensured via single slit diffraction experiment (Fig. 5 inset (a)). The azimuthal phase of the counter propagating LG beams resulting in bent central fringe [41] is not seen in the single slit diffraction pattern. This indicates that both the beams travel through the interferometer along exactly the same path, but in opposite directions and a bright fringe falls on the phase singularity at the beam center (Fig. 5 (a)). Within paraxial-approximation, we can assume that the output beam has plane wavefront due to coherent superposition between the two counter-propagating beams. Changing the phase-difference between the beams by $\pi$, a dark fringe can be made to fall on the phase singularity. This can be realized by adjusting either the geometric phase (input QWP-HWP) or the dynamic phase (mirror M$_2$) between the interfering beams. Now, one of the counter-propagating LG beams can be selectively suppressed by inserting an oriented polarizer in the interferometer arm. If the polarizer is kept within the interferometer, oriented at 90$^o$, the output horizontal polarized $LG_0^{+1}$ beam shows a continuously bent (left-to-right) central bright fringe due to the azimuthal phase of the beam of topological charge +1 (Fig. 5 (b)). Rotating the polarizer to 0$^o$, the output vertical polarized $LG_0^{-1}$ beam shows continuously bent central bright fringe in a direction opposite to the previous one, due to the LG beam's topological charge of -1 (Fig. 5 (c)). As mentioned, by simply changing the phase-difference between the beams by $\pi$, the tilted central fringe can be made to coincide with a dark fringe instead. For a 45$^o$ orientation of the polarizer, both the modes are present in the output beam and non-maximal non-separable superposition of the modes leads to the diffraction fringe shown in Fig. 5 (a), but with less intensity.

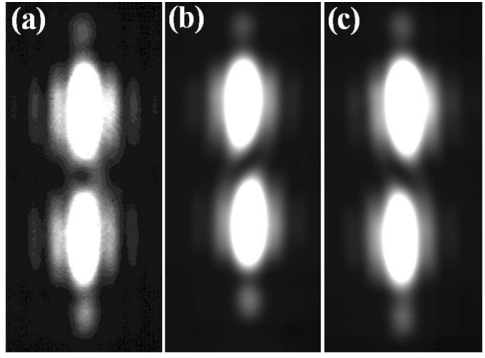

FIG. 5. CCD images of the single slit diffraction patterns measured for (a) non-separable beam from the PSI with plane wavefront, (b) horizontal polarized $LG_0^{+1}$ and (c) vertical polarized $LG_0^{-1}$ beams with helical wavefronts of opposite topological charges.

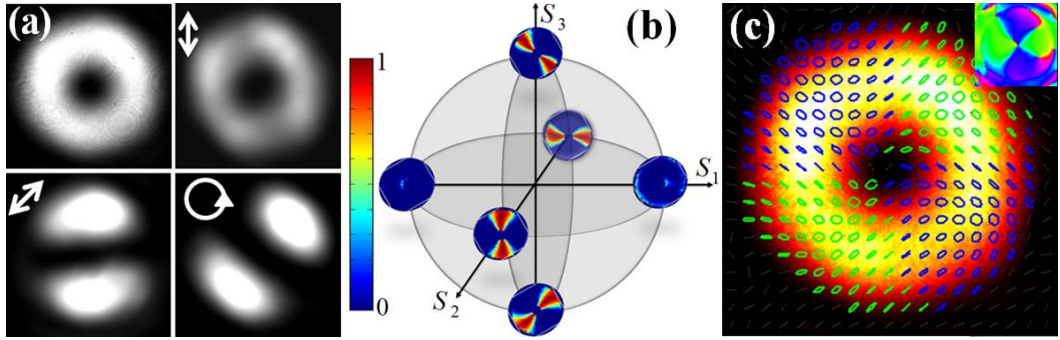

FIG. 6. (Color online) (a) Output beam from the PSI and for different polarization projections (indicated by white arrows); (b) Normalized Stokes parameters of the measured $\Phi^+$ Bell state mapped on the Poincaré sphere; (c) State of polarization calculated from measurements is plotted over the intensity image and corresponding polarization ellipse orientation is shown in inset.

As the superposed beams are orthogonally polarized, there is no spatial interference pattern observed on the screen. However, the resulting polarization interference shows up as azimuthal variation of the modal intensity. In Figure 6 (a) we show the direct CCD image output from the interferometer and the images obtained for different polarization projections: for vertical and horizontal polarization projections, the output beam is $LG_0^{\pm 1}$ modes which for diagonal-anti-diagonal and RCP-LCP projections are Hermite-Gaussian (HG) modes with different orientations. These observations that the input $|H\rangle$ and $|V\rangle$ polarized $LG_0^{\pm 1}$ modes acquiring complete projections in other orthogonal states of polarization (D-A; R-L) indicates high degree of correlation between the two DoF of the polarized input modes. The Stokes parameters $(S_0 - S_3)$ are then calculated point-wise using the CCD images measured for 6 different QWP-Polarizer orientations (Fig. 4 (b)) of the output beam, using the method discussed in Ref. [42]. The measurements are shown on the Poincaré sphere of the Stokes parameters in Fig. 6 (b). The polarization-spatial mode non-separability has resulted in an optical beam wherein the overall state of polarization in the beam cross-section is inhomogeneous as shown in the composite

Figure 6 (c); wherein the SoP is diagonal – anti-diagonal along the horizontal $(\hat{x})$ and vertical $(\hat{y})$ axis, and is RCP – LCP along the direction with 45° rotated axis. This confirms that we have generated an optical beam, corresponding to the Bell state $\Phi^+$ (Fig. 2 (d)), wherein the polarization and spatial mode DoF are highly correlated.

As a witness to having achieved non-separability between the two binary DoF, also referred to as hybrid entanglement in classical beams or shortly classical entanglement we arrange the setup shown in Fig. 4 (c) to perform quantum-inspired Bell state projection measurement in the polarization domain. The output SIP beam from the PSI passes through a 50-50 beamsplitter (BS). The HWP in the transmission arm is kept oriented at 22.5° so that the transmitted (D$_1$) and reflected (D$_2$) beams at the polarizing beamsplitter PBS$_1$ correspond to the ±45° projections of the input beam representing the diagonal and anti-diagonal polarized Hermite-Gaussian $(HG_{01(10)})$ modes respectively, the $|\hat{e}_D \psi_{01}\rangle$ and $|\hat{e}_A \psi_{10}\rangle$ modes shown in Fig. 2 (c). The reflected beam from the BS pass through the QWP oriented at 45° which results in the left and right circular polarization projections at PBS$_2$ corresponding to the $|\hat{e}_L \psi_d\rangle$ and $|\hat{e}_R \psi_a\rangle$ modes shown in Fig. 2 (c).

Next, we investigate the CHSH form of Bell's inequality violation [44] as witness to the non-separability between polarization and spatial mode DoF of the SIP beam. The Bell parameter B is defined as [4],

$$B = E(\theta_1, \theta_2) - E(\theta_1, \theta_2') + E(\theta_1', \theta_2) + E(\theta_1', \theta_2') \tag{11}$$

Where $E(\theta_1, \theta_2)$ are calculated by measuring the intensity $I(\theta_1, \theta_2)$ at a fixed point on the beam cross section using the CCD camera as

$$E(\theta_1, \theta_2) = \frac{I(\theta_1, \theta_2) + I(\theta_1 + \frac{\pi}{4}, \theta_2 + \frac{\pi}{4}) - I(\theta_1 + \frac{\pi}{4}, \theta_2) - I(\theta_1, \theta_2 + \frac{\pi}{4})}{I(\theta_1, \theta_2) + I(\theta_1 + \frac{\pi}{4}, \theta_2 + \frac{\pi}{4}) + I(\theta_1 + \frac{\pi}{4}, \theta_2) + I(\theta_1, \theta_2 + \frac{\pi}{4})} \tag{12}$$

Here $\theta_1$ and $\theta_2$ are the orientation angles of the Dove prism (DP) and the HWP respectively. For the orientation angles of $\theta_1 = 0$, $\theta_2 = 0$, $\theta_1' = \pi/8$ and $\theta_2' = \pi/8$ we use eqns. (11) and (12) to calculate the Bell parameter B = 2.33 ± 0.024 at the point marked in red on the beam cross section shown in Fig. 7.

Then the degree of non-separability between the polarization and spatial mode DoF is obtained via concurrence measurement using the setup shown in Fig. 4 (d). The output beam from the interferometer pass through the QWP kept oriented at 45° to convert the $\Phi^+$ SIP beam into one of the CVBs. The Dove prism and HWP mounted on rotation stages allow us to independently access the spatial mode and SoP degrees of freedom of the beam. A polarizer is kept oriented at 0° enables measurement of spatial mode projections using the CCD camera. Fixing the DP orientation and rotating the HWP results in the rotation of the spatial mode pattern of the beam which can be mapped on the spatial mode sphere corresponding to the S$_1$-S$_2$ plane (Fig. 1). For a fixed angle of the DP of $\theta_1 = 0°$ the HWP is rotated from $\theta_2 = 0° - 180°$ and the CCD images of the mode obtained as a function of HWP rotation angle are shown in the top row in Fig. 7. It can be seen that the two-lobe Hermite-Gaussian (HG) mode pattern rotates as a function of $\theta_2$, indicating clearly that changing the SoP of the non-separable SIP beam changes its modal characteristics. A square box of size 10x10 pixels, at a fixed position on the CCD camera, (indicated by red box in Fig. 7) is used to obtain the average intensity of the beam for different $\theta_2$ values and is used to plot the graph shown by black squares and fitted to $\cos^2(\theta_1 - \theta_2)$ function (Fig. 7). Now changing the orientation angle of the DP to $\theta_1 = 22.5°$ changes the mode orientation projected by the polarizer. Such behaviour is again a clear cut indication that the polarization and spatial mode DoF of the SIP beam are non-separable. The behaviour of the spatial mode as a function of the DP angle for a fixed HWP angle is shown on the right side column of Fig. 7. From all the images obtained for different $(\theta_1, \theta_2)$ values we calculate the average intensity at a fixed point and plot the concurrence measurement graph shown in Fig. 7. From the sinusoidal intensity variations shown in the graph, the concurrence is calculated to be C = 0.912 ± 0.027, clearly confirming high degree of non-separability between the polarization and spatial mode DoF resulting in the SIP beam from the interferometer. Following the same procedure we can obtain the concurrence for any point on the beam cross-section. Note: small intensity variations between the curves are attributed to weak polarization dependency of the Dove prism [43].

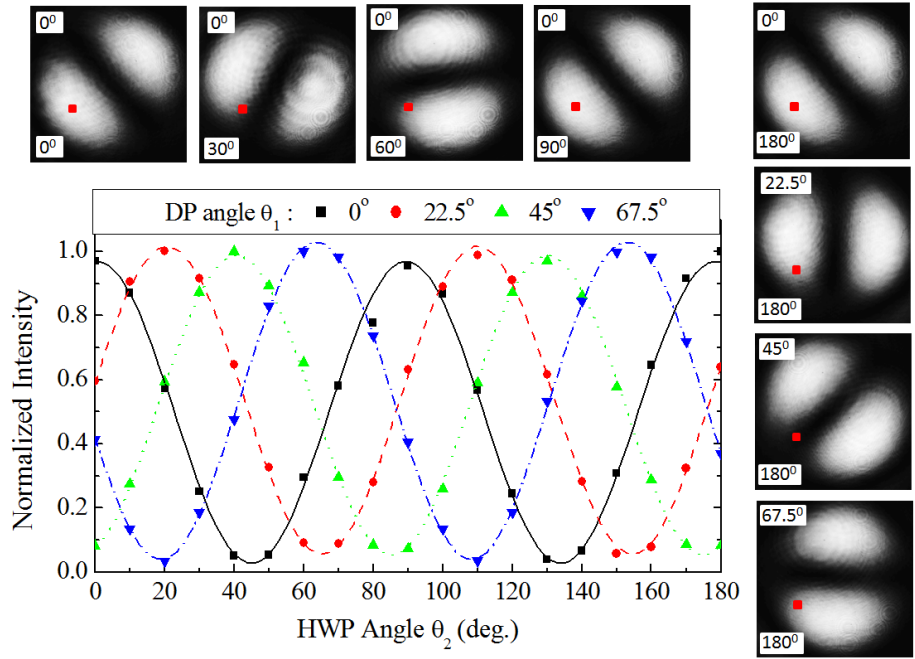

FIG. 7. (Color online) Concurrence measurement carried out using the setup shown in Fig. 4 (d). Top row of intensity images correspond to HWP rotation angle ($\theta_1$ = 0°, 30°, 60° and 90°) for a fixed angle of the Dove prism ($\theta_2$ = 0°) and images shown on the right side correspond to a fixed HWP angle of $\theta_1$ = 180° and different DP angle ($\theta_2$ = 0°, 22.5°, 45° and 67.5°). Red box of size (10x10 pixels) in the images correspond to the point where the intensity values are considered in all the images taken for different ($\theta_1, \theta_2$). The normalized intensity values are shown as a function of HWP angle for different fixed DP angle. The average concurrence value calculated from all the measurements is C = 0.912 ± 0.027. Small intensity variations between the curves are attributed to weak polarization dependency of the Dove prism.

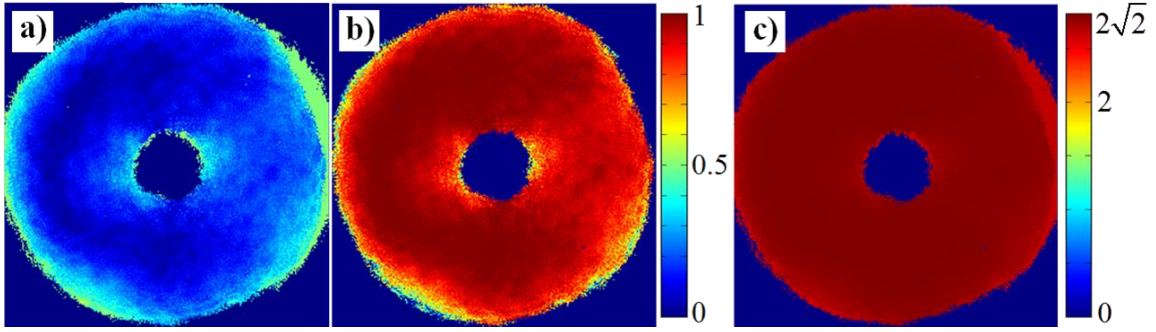

FIG. 8. (Color online) (a) Degree of polarization ($D_{pol}$), (b) linear entropy ($S_L$) and (c) maximized Bell's measure ($B_{max}$) for the hybrid entangled SIP beam shown in Fig. 4 (c) calculated using the imaging Stokes polarization measurements and eqns. (4), (9) and (6).

After these initial quantum-inspired entanglement witness and quantification measurements for the SIP beam we use the formalism of using reduced polarization coherency matrix to quantify the non-separability between the two binary DoF. From the experimentally measured point-wise Stokes parameters, we construct the reduced polarization-coherency matrix given by eqn. (7) at each point on the beam cross-section. In the context of the optical beam wherein the SoP shows azimuthal variation, the reduced $2\times 2$ coherency matrix characterises the polarization DoF (eqn. (4)), with $r_1$ and $r_2$ corresponding to the positions of the two pinholes [8]. This is $\mathbf{G}_{pol}$ describing the partial polarization of the beam-field, after integrating over the modal space. As mentioned in Ref. 8, determining $\mathbf{G}_{pol}$ requires a detector and polarization components with no spatial discrimination, covering both pinholes, which in our case is performed inherently pixel-wise by the CCD camera. The Stokes polarimetry measurements gives us point-wise modal information in the beam cross-section. Considering the reduced polarization coherency matrix given in eqn. (4), the diagonal elements give intensity averages associated with $E_x$ and $E_y$ components of the electric field and off-diagonal elements give the cross-correlation between them. As discussed earlier we choose the two points $r_1$ and $r_2$ in the beam cross-section with an angular separation of $90°$ to maximize the entropy. Using these, the degree of polarization ($D_{pol}$) of the maximally entangled beam that belongs to the $\Phi^+$ Bell-state is calculated using eqn. (6). Figure 8 (a) shows the calculated $D_{pol}$ in the beam cross-section. Azimuthally averaging across 360x80 pixels (ignoring the inner and outer edges of the beam) gives an average value of $D_{pol} = 0.117 \pm 0.057$. The experimentally obtained higher than theoretically expected value of $D_{pol} = |\cos 2\theta| = 0$ for $\theta = 45°$ [8] is possibly due to non-ideal experimental conditions and polarization components used in the measurements and can be improved to match the simulation results shown in Fig. 9. It is important to note here that the small $D_{pol}$ value is due to incoherence in the beam that appears due to non-separable polarization-spatial mode entanglement [8]. To ensure that the two DoF of the beam are maximally entangled across the entire beam cross-section, we calculate the linear entropy $S_L$ using eqn. (8) and the result obtained is shown in Fig. 8 (b). The average value of $S_L = 0.968 \pm 0.031$ for the $\Phi^+$ Bell-state beam generated using the PSI emphasizes high degree of hybrid entanglement achieved. The maximized Bell's measure calculated using eqn. (9) is shown in Fig. 8 (c) and the spatially averaged value of $B_{max}$ is 2.824

± 0.004, which shows a bound of $B_{max} \leq 2\sqrt{2}$ [8] in the beam cross-section, further confirming the high degree of entanglement of the SIP beam. For optical beam with separable DoF, corresponding to homogeneously polarized (scalar) beam with uniform SoP in the beam cross-section, the entanglement parameters are calculated from the Stokes parameters to be $D_{pol} = 0.997 \pm 0.003$; $S_L = 0.176 \pm 0.059$ and $B_{max} = 2.007 \pm 0.006$.

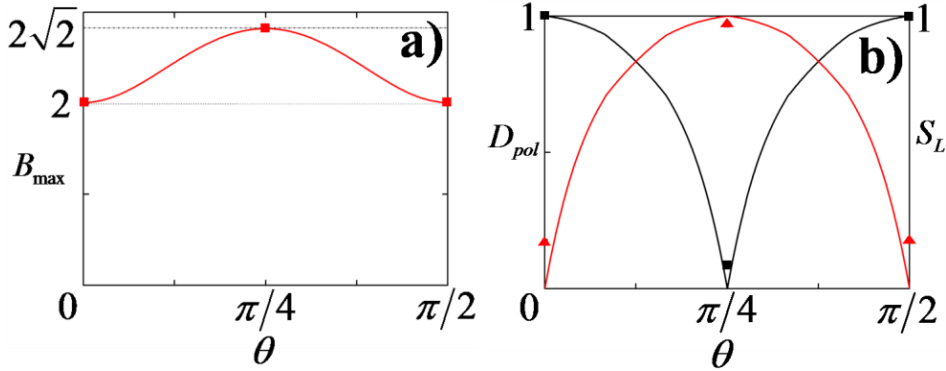

FIG. 9. (Color online). Simulated results show variations in entanglement parameters $D_{pol}, S_L, B_{max}$ as a function of polarizer angle $\theta$. Experimental results are shown in symbols (■: $D_{pol}$, ▲: $S_L$ in (b)), statistical errors are less than the symbol size. As can be seen, maximum entanglement is achieved for $\theta = \pi/4$ and the beams are separable for $\theta = 0$ and $\pi/2$.

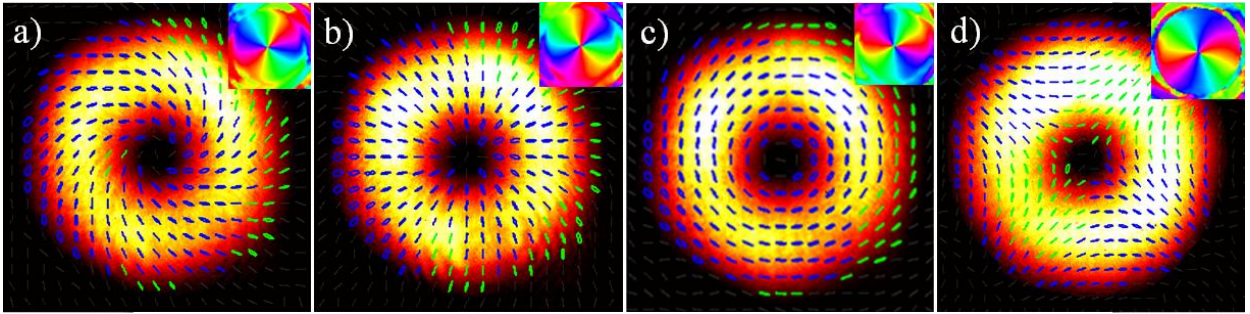

FIG. 10. (Color online) The hybrid entangled beam corresponding to $\Phi^+$ Bell-state generated at the PSI output is passed through Q-H wave plate combination oriented at (a) (45°-0°), (b) (45°-22.5°), (c) (45°-67.5°) and (45°–Nil) generate entangled beams respectively with spiral, radial, azimuthal and hybrid polarization variations in the beam cross-section. Inset: corresponding polarization ellipse orientation.

Further, to see variations in the entanglement parameters of the output beam from maximally entangled to separable state, the Bell state of the SIP beam can be written as $\Phi^+ = \sin\theta |H\rangle\psi_l + e^{i\varphi}\cos\theta |V\rangle\psi_r$, where $\theta$ is the angle the polarizer inserted within the interferometer, makes with vertical. Now, by rotating the polarizer from $\theta = 0°$ to $90°$ the Stokes parameters can be measured to calculate $D_{pol}, S_L, B_{max}$ as discussed above. Figure 9 shows the simulation of the variations in the entanglement parameters obtained as a function of $\theta$ and the corresponding experimentally obtained values. The SIP beam is maximally non-separable for $\theta_1 = 45°$ and is separable for $\theta_1 = 0$ and $90°$. The statistical errors in the entanglement parameter values are smaller than the size of the symbols used.

Finally, passing the $\Phi^+$ Bell state beam generated using the PSI through a QWP-HWP combination (Fig. 4 inset (e)), we successfully convert the generic SIP beam into different standard CVBs with spiral, radial, azimuthal and hybrid polarization variations in the beam cross section, for different orientations of the waveplates as shown in Fig. 10 (a)-(d), without affecting the overall degree of entanglement. Also shown in the inset of Fig. 10 are the polarization ellipse orientations of the cylindrical vector beams with non-separable DoF.

## V. CONCLUSION

Polarization and spatial mode binary DoF of an optical beam are maximally entangled in a polarization Sagnac interferometer to generate SIP beam, represented on a HPS. We show that changing SoP of the beam modifies the spatial mode characteristics and vice-versa, indicating non-separability between the two DoF. Quantum-inspired Bell-state, violation of CHSH form of Bell's inequality (B = 2.33 ± 0.024) and concurrence (C = 0.912 ± 0.027) measurements are carried out on the SIP beam as witness to the non-separability between the DoF. The reduced polarization coherency matrix formalism is used to quantify classical entanglement via point-wise Stokes polarimetry measurements. The angular settings of pairs points in the beam cross-section is spanned to maximize the linear entropy ($S_L$ = 0.968 ± 0.031) and the Bell's measure ($B_{max}$ = 2.824 ± 0.004). Thus polarimetry based quantification of entanglement proves maximal non-separability between the DoF in realizing the SIP beam. Unitary transformation of the SIP beam using lossless waveplates and / or mode converters enables realization of radial, azimuthal, spiral and hybrid type CVBs. Such beams have the potential to enable some implementation of quantum game, quantum information and protocols using classical optical beams.

## ACKNOWLEDGEMENTS

The authors thank Department of Science and Technology (DST), India for funding projects which enabled us to extend to the present work and Rishabh Pandey for critical reading of the manuscript's early versions. CTS thanks UGC-BSR for research fellowship. The authors would like to thank the referee for his comments and for bringing to our notice Ref. [4] which appeared around the same time as our submission.